\documentclass[sigconf]{acmart}
\AtBeginDocument{%
  }

\setcopyright{acmlicensed}
\copyrightyear{2025}
\acmYear{2025}
\acmDOI{}
\acmConference[CUG]{CUG 2025}{May 04--08,
  2025}{New York City, NY}
\acmISBN{}

\begin{document}


\title{FirecREST v2: Lessons Learned from Redesigning an API for Scalable HPC Resource Access\\
}

\author{Elia Palme}
\email{elia.palme@cscs.ch}
\affiliation{%
  \institution{CSCS -- Swiss National Supercomputing Centre}
  \city{Lugano}
  \country{Switzerland}
}
\author{Juan Pablo Dorsch}
\email{juanpablo.dorsch@cscs.ch}
\affiliation{%
  \institution{CSCS -- Swiss National Supercomputing Centre}
  \city{Lugano}
  \country{Switzerland}
}
\author{Ali Khosravi}
\email{ali.khosravi@psi.ch}
\author{Giovanni Pizzi}
\email{giovanni.pizzi@psi.ch}
\affiliation{%
  \institution{PSI Center for Scientific Computing, Theory, and Data}
  \city{5232 Villigen PSI}
  \country{Switzerland}
}
\author{Francesco Pagnamenta}
\author{Andrea Ceriani}
\author{Eirini Koutsaniti}
\author{Rafael Sarmiento}
\author{Ivano Bonesana}
\author{Alejandro Dabin}
\affiliation{%
  \institution{CSCS -- Swiss National Supercomputing Centre}
  \city{Lugano}
  \country{Switzerland}
}

\renewcommand{\shortauthors}{Palme et al.}

\begin{abstract}
Introducing FirecREST v2, the next generation of our open-source RESTful API for programmatic access to HPC resources. FirecREST v2 delivers a $\sim$100x performance improvement over its predecessor. This paper explores the lessons learned from redesigning FirecREST from the ground up, with a focus on integrating enhanced security and high throughput as core requirements.
We provide a detailed account of our systematic performance testing methodology, highlighting common bottlenecks in proxy-based APIs with intensive I/O operations. Key design and architectural changes that enabled these performance gains are presented. Finally, we demonstrate the impact of these improvements, supported by independent peer validation, and discuss opportunities for further improvements.
\end{abstract}


\maketitle

\section{Introduction}
The HPC industry has grown increasingly competitive with the entry of new players, such as cloud solution providers. User expectations continue to rise, requiring us to adapt our services to meet these higher demands. One such service is the FirecREST API\cite{f7t}\cite{f7t-cug2024}, a RESTful web API written in Python for programmatic HPC resource access. FirecREST is a core offering of the Swiss National Supercomputing Center (CSCS). The API supports a wide range of use cases, from enabling pipelines to integrating workflow engines like AiiDA\cite{AiiDA}.
Over time, the growing demands from users and partners leveraging FirecREST prompted us to reassess its performance and security features. Through extensive performance analysis, we identified key bottlenecks in the API, underscoring the need for a complete redesign of FirecREST from the ground up. On the security side, in addition to authentication, a new authorization feature has been introduced.   

\section{Performance Measurement}
To evaluate the performance of the FirecREST API under realistic load conditions, we developed a specific testing method. Since the overall performance of a proxy server is inherently tied to the performance of its underlying services, accurate measurement requires two key conditions: (A) a representative set of requests (e.g., filesystem access, job scheduling, data transfer, etc.) and (B) forwarding calls to an HPC cluster with response times that accurately reflect real-world conditions.

To address both (A) and (B), we established a protocol for conducting reproducible stress tests in a production-like environment.

\subsection{Load Simulation}

To simulate realistic load conditions, we utilize API testing tools such as Postman\cite{postman} to design and execute workflows based on real-world use cases.

\begin{figure}[ht]
     \centering
     \includegraphics[width=\linewidth]{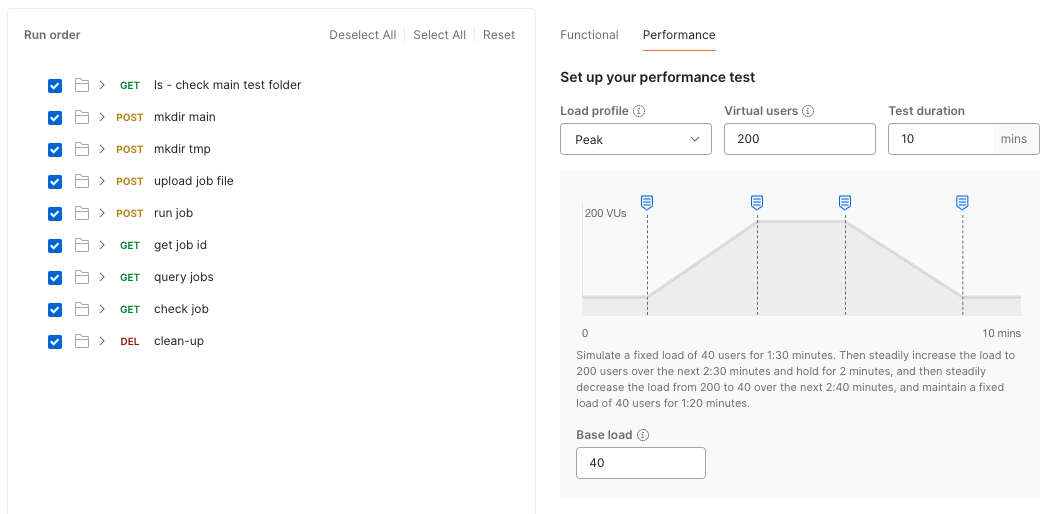}
     \caption{Postman Performance Testing UI - screenshot.}
     \label{fig:postman-screenshot}
     \Description{A screenshot of the Postman user interface.}
 \end{figure}

As shown in the screenshot (\ref{fig:postman-screenshot}), Postman is used to simulate load by executing a series of sequential requests that replicate a typical workload. Multiple sets of sequential requests are run concurrently by a specified number of clients over a defined time period. Additionally, the configuration allows for dynamically increasing the number of clients over time to model traffic peaks effectively.
\subsection{Test environment}
To ensure that the performance results are representative of real-world usage, load testing is conducted using a FirecREST instance directly connected to our staging environment. This staging environment consists of a small HPC cluster that is part of the CSCS Alps infrastructure—an HPE Cray EX system featuring heterogeneous hardware, including AMD and Nvidia CPUs and GPUs.

The test workload was carefully designed by analyzing typical user interactions with FirecREST. It consists of a structured sequence of API calls that simulate a complete job lifecycle, including file system setup, job submission, execution monitoring, and output retrieval. This workload is implemented in Postman as a sequential series of API requests, where each request triggers specific FirecREST end-points under controlled conditions.

To ensure the reliability and accuracy of the test, Postman validates each API response against predefined assertions. If a response meets all validation criteria, the test flow proceeds to the next step. Otherwise, the execution is halted. This validation process is crucial for identifying the system’s performance limits without quality degradation.

Furthermore, by incrementally increasing the load, we can assess FirecREST’s scalability, response times, and stability under different stress levels. These insights help identifying performance constrains and bottlenecks. 

\section{Performance Constraints}
Our performance measurement tests revealed several bottlenecks in processing the incoming FirecREST requests.

To pinpoint the root causes of these bottlenecks, we employed multiple analytical methods:
\begin{itemize}
\item Log Analysis: By tracking and comparing timestamps logged for individual API calls, we measured the frequency and response time of FirecREST API requests. This analysis was particularly useful in uncovering limitations in Python multithreading and highlighting how Gunicorn\cite{gunicorn}, the HTTP server used in FirecREST, sequentially processed incoming requests, thereby creating execution delays under high loads.
\item Network Traffic Analysis: We examined FirecREST API network activity to identify performance-impacting downstream services. This investigation revealed that certain operations, such as JWT token introspection\footnote{In contrast to JWT token local validation the token introspection implies to sends the token to the identity provider’s introspection endpoint to determine its validity} , were invoked frequently and significantly contributed to response time overhead.
\item Resource Usage Analysis: By monitoring the back-end system resource consumption (primarily CPU and memory) we identified operations with high computational overhead. A key finding was that establishing SSH connections imposes a substantial resource burden, impacting overall service responsiveness.
\end{itemize}
These in-depth performance assessments uncovered three primary bottlenecks that directly influenced FirecREST’s efficiency. The insights gained from this analysis guided the architectural redesign of FirecREST v2, specifically addressing these limitations to enhance scalability, concurrency, and overall system performance.

\subsection{Multi-threaded VS asynchronous Web Server}
The FirecREST API mostly acts as a proxy, its workload primarily involves processing requests and forwarding them, with the forwarding step typically being an I/O blocking operation. Python offers two main approaches to handle concurrent I/O blocking requests: multi-threading and asyncio\cite{asyncio}.

In the multi-threading approach, each request is handled by a separate thread. When a thread is blocked while waiting for a forwarded request’s response, the Python interpreter reallocates compute resources to other threads. However, this approach has scalability limitations due to the maximum number of threads the server is allowed to spawn. Additionally, each thread introduces a small memory overhead.

Note about Python: Before Python 3.13, effective multi-threading was heavily limited by the Global Interpreter Lock (GIL), which largely restricted threads to sequential execution. Starting with Python 3.13, free-threading has been introduced as an experimental feature. However, it is not enabled by default and may not be supported by all Python libraries.

In contrast, Python’s asyncio offers better scalability as it is not constrained by an upper limit on concurrent asynchronous operations. All requests are handled within a single thread, with the asyncio event loop efficiently managing tasks by switching between them while they await for I/O.

Our benchmarks (see Section \ref{section:benchmarking}) demonstrated that Python-based asyncio web servers, such as Uvicorn\cite{uvicorn}, which power FirecREST v2, significantly outperformed the multi-threaded alternatives, such as Gunicorn\cite{gunicorn}, used in FirecREST v1. The performance gains stem from the asynchronous driven architecture of Uvicorn, which efficiently handles I/O constrained workloads with lower overhead compared to traditional thread-based servers. This results in reduced latency, improved resource utilization, and better scalability under heavy load.

\subsection{Offline VS Online Authentication}
The FirecREST authentication layer implements the OpenID Connect (OIDC)\cite{oidc} standard, requiring incoming requests to include a JSON Web Token (JWT)\cite{jwt} issued by a trusted identity provider. The OIDC standard supports both offline (local signature validation) and online (introspection) JWT token validation mechanisms.

Online validation involves forwarding the token to the identity provider for verification, which introduces additional network requests and latency. In contrast, offline validation verifies the token’s signature locally using the identity provider’s public certificates.

The main drawbacks of using offiline validation is the inability to revoke tokens after their have been granted and the data staleness (e.g. user role changes not being reflected). However, for FirecREST, these concerns are mitigated by employing very short-lived tokens. The brief lifespan of these tokens reduces the window of opportunity for exploitation, making immediate revocation less critical and minimizing the impact of potential replay attacks or stale data issues.

\subsection{SSH Authentication overhead}
Some operations, such as listing remote files, require FirecREST to open an SSH channel. Our tests revealed that beyond a certain threshold, the SSH daemon struggles to handle a high volume of concurrent connection requests, leading to queued requests. While the actual threshold depends on the amount of resources dedicated to the SSH daemon, we found that the service was difficult to scale by adding resources. The bottom line is that SSH is not meant for high-throughput.

SSH daemons, like OpenSSH, have inherent limitations in handling a large number of concurrent connections. By default, the MaxStartups parameter in the sshd\_config file limits the number of concurrent unauthenticated connections to 10. This means that if more than 10 unauthenticated connections are attempted simultaneously, additional connection attempts may be dropped or delayed. While this value can be increased, doing so may lead to increased resource consumption and potential instability.

Similarly, the MaxSessions parameter controls the maximum number of open sessions permitted per network connection, with a default value of 10. Adjusting this parameter allows more sessions per connection but does not address the overall scalability challenges when numerous clients initiate separate SSH connections.

The underlying architecture of SSH, designed primarily for secure remote administration, is not optimized for high-throughput scenarios involving thousands of simultaneous connections.

In our experience with FirecREST, even after tuning the SSH daemon configuration and allocating additional resources, we observed diminishing returns in performance improvements.

To address this limitation, we implemented an optimization that leverages the fact that, under high-throughput conditions, resource access requests are often made by the same user. As a result, multiple requests can share the same SSH connection. FirecREST’s SSH connection pool helps alleviate the load on the SSH daemon by reusing existing connections for requests from the same user, significantly improving scalability.

Our benchmarks (see Section \ref{section:benchmarking}) showed that introducing a SSH connections pool significantly improved the performance under high-throughput regime.

\section{Improved Firecrest v2 Architecture}
Based on our performance analysis, we undertook a comprehensive redesign of the FirecREST API to address identified constraints and bottlenecks. Our goal was to develop a more efficient and scalable API by adhering to the following core principles:

\begin{itemize}
\item Adopt Broadly Used Standards: Utilize widely adopted protocols and conventions to ensure interoperability and facilitate integration.
\item Abstract Underlying Services: Create a uniform interface that encapsulates the complexities of various backend services, facilitating seamless interaction across different vendor systems (e.g. Slurm and OpenPBS).
\item Minimize Interactions with Underlying Services: Reduce the number of direct communications with backend systems to decrease latency and improve overall performance.
\item Implement Asynchronous I/O Operations: Design all input/output processes to operate asynchronously, enhancing responsiveness and scalability.
\item Ensure Statelessness: Avoid maintaining session states or persisting user-specific information between requests to promote scalability and simplify horizontal scalability.
\end{itemize}

\begin{figure}[ht]
     \centering
     \includegraphics[width=\linewidth]{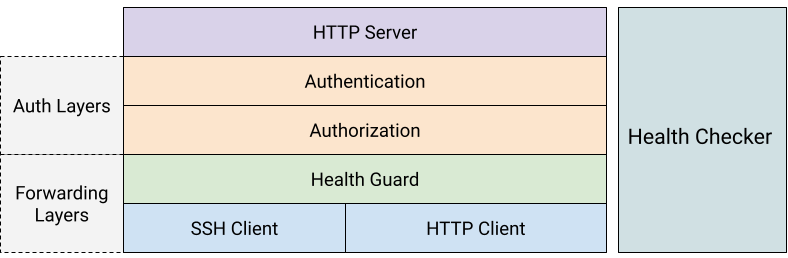}
     \caption{FirecREST API layered architecture.}
     \label{fig:architecture}
     \Description{A diagram that shows FirecREST layered architecture.}
 \end{figure}

Guided by these principles, we envisioned the new version of FirecREST as a lightweight proxy with a layered architecture, providing a RESTful web interface. Incoming requests traverse through four distinct layers (see Figure \ref{fig:architecture}), each performing specific checks before forwarding the request to the appropriate subsystem. Additionally, a local task manager executes various housekeeping processes to monitor the state of the underlying services. The status information from these checks informs the forwarding layers, enabling them to determine the safety and appropriateness of forwarding each request.

The API proxy layers are organized into two primary groups:

\begin{itemize}
\item Authentication and Authorization Layers: These initial layers authenticate the user and verify authorization to access the requested resource.
\item Request Forwarding Layers: Upon successful authentication and authorization, these layers assess the readiness of the requested resource to handle new requests and determine the optimal communication channel (e.g., SSH, HTTP) for forwarding.
\end{itemize}

This architecture aligns with best practices for RESTful API design, emphasizing stateless interactions, standardized interfaces, and scalability.  By implementing asynchronous operations and minimizing direct interactions with underlying services, the redesigned FirecREST API enhances performance and responsiveness.

\subsection{HTTP Server, framework and libraries}
As part of the overall redesign of the FirecREST API, we meticulously selected frameworks and libraries that align with our core principles of efficiency, scalability, and adherence to widely accepted standards. The revamped software stack is engineered to support end-to-end asynchronous operations, comprising the following core components:
\begin{itemize}
\item Uvicorn\cite{uvicorn}: A high-performance ASGI (Asynchronous Server Gateway Interface) server. Uvicorn is built upon uvloop, a fast event loop implemented in Cython, which significantly enhances the speed of asynchronous operations. It also utilizes httptools for HTTP protocol handling, ensuring efficient request parsing and response generation.
\item FastAPI\cite{fastapi}: A modern, high-performance web framework for building APIs via dedicated Python decorators. It natively supports asynchronous programming, allowing for the development of endpoints that can handle concurrent requests seamlessly.
\item Asyncio: Python’s built-in library for writing concurrent code using the async and await syntax. Asyncio provides the framework for asynchronous programming, enabling the development of code that can execute non-blocking calls and manage multiple tasks concurrently within a single thread. This is particularly beneficial for I/O-bound operations, as it allows the application to remain responsive while waiting for external events.
\item AsyncSSH\cite{asyncssh}: An asynchronous SSH client built on top of Asyncio. AsyncSSH enables the FirecREST API to establish secure SSH connections asynchronously, facilitating non-blocking interactions with the underlying services. This is crucial for operations that require communication with HPC resources over SSH, as it ensures that such interactions do not impede the responsiveness of the API.
\end{itemize}

By integrating these components, the FirecREST API achieves a fully asynchronous architecture that enhances its ability to efficiently process high requests volume. This design not only improves the API’s responsiveness but also aligns with modern web development practices, ensuring compatibility with a wide range of client applications and services.

\subsection{Auth Layers}
The authentication layers within FirecREST are responsible for both authenticating and authorizing requests to access the desired HPC resources. All requests, except for diagnostic ones, must pass through these layers, undergoing authentication followed by authorization.

\subsubsection{Authentication}
The authentication layer relies on the OpenID Connect (OIDC)\cite{oidc} standard, ensuring compatibility with various authentication flows. However, given the utilization context of the API, the Client Credentials and Authorization Code flows are particularly pertinent.

The authentication process begins with the inspection of the JSON Web Token (JWT)\cite{jwt} accompanying each request. Such token needs to be passed via the HTTP Authentication header: 

\begin{verbatim*}
Authorization: Bearer {token}
\end{verbatim*}

The JWT must be signed by a trusted identity provider and include claims indicating the user or service account making the request. In alignment with our design principles, FirecREST performs offline JWT token validation, reducing interactions with the identity provider and thereby decreasing request handling latency.

FirecREST can be configured with multiple OIDC certificate URLs, enabling support for multiple identity providers and multiple realms per identity provider. At startup, FirecREST downloads all OIDC certificates and creates a map of key IDs (kid) to their respective public keys. To validate the signature of an incoming JWT, FirecREST accesses the unverified token header to extract the kid claim, which is then used to find the appropriate public key for signature verification.

Upon successful token signature validation, additional token claims are extracted. The preferred\_username or username claim are mandatory. These claims are used to map the incoming request to a POSIX user on the HPC system. FirecREST requires that all incoming requests can be mapped to a user. OIDC flows that are not bound to a user (e.g., Client Credentials) will require the identity provider to add a custom mapping to assign a service account to the client, enabling FirecREST to determine the appropriate POSIX user for accessing the HPC resources.

\subsubsection{Authorization}
Following authentication, the authorization layer determines whether the authenticated user or service account possesses the necessary permissions to access the requested HPC resource. Given that a single FirecREST instance can interface with multiple HPC clusters, the authorization layer ensures that a valid access relationship exists between the user and the specific HPC system. This relationship can be established through claims embedded in the JSON Web Token (JWT) or by leveraging an external access control authorization service.

The JWT claim approach is straightforward and efficient. In this method, the identity provider includes a claim in the token that contains a list of HPC system names to which the user has access. Upon processing the request, FirecREST verifies that the requested system is present in the user’s token claims. While this method offers simplicity and low latency, it has limitations, notably the size constraints of JWTs, which restrict the number of authorized services that can be listed. This constraint is not imposed by the JWT itself but by the fact that it is transmitted over HTTP. Some Web servers impose a limit of 8 KB for the total size of all headers in a request. Since the JWT is included in the authorization header, it is affected by such a limit.

Alternatively, FirecREST can utilize an external OIDC\cite{oidc} compatible authorization service to perform fine-grained access control checks. Authorization service such as OpenFGA\cite{openfga} allow developers to define access control policies using a relationship-based access control model.

In this approach, FirecREST queries OpenFGA to verify if the user has the appropriate relationship with the requested HPC system. OpenFGA’s modeling language facilitates the definition of various authorization paradigms, including Role-Based Access Control (RBAC) and Attribute-Based Access Control (ABAC). By externalizing the authorization logic, OpenFGA enables centralized management of access policies, enhancing security and compliance.

\subsection{Forwarding Layers}
The Forwarding Layers in FirecREST are designed to abstract and manage interactions with various HPC subsystems, such as job schedulers, filesystems, and storage systems. Its modular architecture employs dedicated clients for each subsystem type (e.g., Slurm, OpenPBS, S3).

\subsubsection{Modular Design and Client Organization}

Clients within the Forwarding Layers are organized by service category (e.g., Schedulers), with each client category adhering to a standardized interface. This design enforces a consistent abstraction across resources of the same type, regardless of the vendor. Consequently, integrating a new vendor involves implementing a new client class specific to that service, enhancing the system’s extensibility and maintainability.

\subsubsection{Communication Protocols}
Furthermore the Forwarding Layers are responsible for routing incoming requests to the appropriate underlying systems using the correct communication protocols, such as SSH or HTTP. This routing ensures that interactions with HPC resources are conducted efficiently and securely. It's each client implementation responsibility to select the appropriate communication protocol.

\subsubsection{SSH Connection Pooling Optimization}

A significant optimization in the FirecREST redesign is the implementation of an SSH connection pool. Establishing a new SSH connection for each new user request incurs considerable overhead. To mitigate this, FirecREST maintains persistent SSH connections for a predefined duration, allowing multiple requests from the same user to share an existing connection. When a new request is received and an active SSH connection is available, a new channel is spawned on the existing connection to handle the request. This approach reduces latency and resource consumption.

The pooling mechanism is facilitated by AsyncSSH\cite{asyncssh}, an asynchronous SSH client library for Python. AsyncSSH supports multiple simultaneous sessions over a single SSH connection, enabling efficient management of SSH channels and connections.

\subsubsection{Healthy Forwarding}
FirecREST incorporates health monitoring to assess the status of target resources. Requests are only forwarded if the health state of the requested resource is known and deemed healthy. This proactive measure prevents overloading backends and issuing requests that might time out, thereby enhancing system reliability.

\subsection{Health Checker}
The Health Checker is an integral component of the FirecREST API, designed to asynchronously monitor the health and status of all integrated subsystems, including the filesystems, the scheduler, and storage solutions. By periodically performing sanity checks, it ensures that each service operates as expected.

\subsubsection{Subsystem-Specific Health Checks}

For each type of service, a dedicated health check implements the logic to determine if the service is healthy. Each health check can be parameterized to define specific monitoring criteria and thresholds, allowing system administrators to tailor the monitoring to the unique characteristics of their HPC environment. If needed health checks can be completely disabled, offering flexibility in monitoring configurations.

\subsubsection{Integration with Forwarding Layers}

The health status information collected by the Health Checker is made available to users for monitoring purposes and is also utilized by the Forwarding Layers. Before forwarding any request, the Forwarding Layers consults the health status to ascertain the readiness of the requested subsystem. If a subsystem is deemed unhealthy or its status is unknown, FirecREST refrains from forwarding requests to prevent potential failures or additional strain on compromised services.

\subsubsection{Performance Optimization through Cached Health Data}

A significant enhancement in the redesigned FirecREST architecture is the caching of health information. By maintaining and referencing cached health statuses, FirecREST reduces the need for real-time health verifications prior to each request. In contrast, the previous approach involved verifying subsystem health before processing most requests, introducing substantial overhead and delaying response times.

\section{Benchmarking}\label{section:benchmarking}
While Postman offers a robust testing interface to gather insights for the API redesign, evaluating real-world performance under production-level conditions could say the last words. This is especially apparent when multiple users are connecting and using the server concurrently.
To assess this, we called on AiiDA\cite{aiida1}\cite{AiiDA}\cite{aiida3} a third party integrator of FirecREST. AiiDA is a workflow manager that helps scientists to interface with HPC systems, particularly for high-throughput calculations. To address AiiDA's use case, we focused on the most demanding endpoint: I/O-bound operations, to measure the latency associated with each individual file request, rather than focusing on the network speed.
For testing, we generated up to 1000 dummy files on the server and conducted asynchronous ‘GET’ calls, using the `AsyncClient` from Python's `httpx` library. 
To isolate request-handling performance, we standardized tests with small file sizes (1 KB), minimizing overhead due to the specific storage backend (e.g., parallel file system) and enabling repeatable test execution.

We measured the total retrieval time for downloading\footnote{To be specific, we tested the ‘/filesystem/ops/download’ endpoint in FirecREST v2, which is equivalent to ‘/utilities/download’ in v1. These endpoints enable stateless executions lasting under 5 seconds, which typically allows downloading files of up to approximately 5 MB. For larger files, due to the nature of REST APIs, truly stateless operations require a two-stage download/upload process. Where authorization layer and I/O are handled in separate requests. However, for this paper, we focus solely on testing single-request downloads.} N files across all deployed versions of Firecrest on CSCS's Eiger\footnote{Eiger was accessed from an external network location. FirecREST deployment on this HPC was running on a single Kubernetes pod for all tests.} cluster. Each test set was executed a minimum of five times with varying numbers of asynchronous workers, and results (summarized in Table 1) are reported as mean values with standard deviations for fluctuations.

\begin{center}
\begin{tabular}{||c c c c||} 
 \hline
 N files & FirecREST v1 & FirecREST v2 & FirecREST v2 \\ & & & (with connections pool) \\ [0.5ex] 
 \hline\hline
 1 & 1.5 ± 0.1 s & 0.8 ± 0.02 s & 0.4 ± 0.3 s \\ 
 \hline
 10 & 13.7 ± 1.1 s & 2.7 ± 0.05 s & 0.5 ± 0.2 s \\ 
 \hline
 100 & 129.6 ± 1.5 s & 19.5 ± 3.4 s & 1.5 ± 0.7 s \\ 
 \hline
 1000 & \textcolor{red}{too long > 1000 s} & 176.3 ± 4.5 s & 15.5 ± 8.9 s \\ 
 \hline
\end{tabular}
\end{center}

The results suggest impressive improvement in FirecREST V2, in particular in the implementation with connections pool. The relatively large fluctuations are related to server stability at the time of the test. Allocating enough resources should diminish the fluctuation in numbers.

\section{Conclusion}
This paper presents a comprehensive methodology for API performance testing and bottleneck identification, utilizing a real-world use case to demonstrate the process of gathering insights and systematically redesigning an API for enhanced efficiency and scalability. Through performance analysis, we identified critical constraints and distilled a set of core principles that guided the complete overhaul of the FirecREST API. The redesigned architecture, characterized by its modular and layered structure, incorporates key components such as the Auth Layers, the Forwarding Layers, and the Health Checker, each playing a critical role in optimizing API interactions with underlying HPC resources.

The Authentication and Authorization Layers ensure secure and efficient access control by adopting the OpenID Connect (OIDC) standard for authentication and OpenFGA for fine-grained authorization.

The Forwarding Layer’s modular design, with dedicated clients for each subsystem type, ensures abstraction and a standardized API interface across various vendors. The implementation of an SSH connection pool significantly reduces latency by reusing existing connections for multiple requests from the same user. 

The Health Checker asynchronously monitors the subsystems health, providing cached health data for monitoring and to ascertain the readiness of the requested subsystem.

Performance evaluations of the re-engineered FirecREST API indicate substantial improvements in response times, reliability, and scalability. These enhancements have been validated through a peer review, confirming the effectiveness of our redesign approach.

\section{Acknowledgments}
This paper incorporates AI-generated text from grammatical and vocabulary suggestions, as well as from translations of foreign-language text snippets into English.

\end{document}